\documentclass[prb,twocolumn,amssymb,showpacs,superscriptaddress]{revtex4}
\usepackage{graphicx,amsmath,epstopdf,color,ulem}

\begin{document}
\title{Pinball liquid phase from Hund's coupling in frustrated transition metal oxides}
\author{Arnaud Ralko}
\affiliation{Univ. Grenoble Alpes, Inst NEEL, F-38042 Grenoble, France}
\affiliation{CNRS, Inst NEEL, F-38042 Grenoble, France}
\author{Jaime Merino}
\affiliation{Departamento de F\'isica Te\'orica de la Materia Condensada, Condensed Matter Physics Center (IFIMAC)
and Instituto Nicol\'as Cabrera,
Universidad Aut\'onoma de Madrid, Madrid 28049, Spain}
\author{Simone Fratini}
\affiliation{Univ. Grenoble Alpes, Inst NEEL, F-38042 Grenoble, France}
\affiliation{CNRS, Inst NEEL, F-38042 Grenoble, France}
\date{\today}
\begin{abstract}
The interplay of non-local Coulomb repulsion and Hund's coupling in the
$d$-orbital manifold in frustrated triangular lattices is analyzed by a
multiband extended Hubbard model.
We find a rich phase diagram  with several competing phases, including a robust
pinball liquid phase, which is an unconventional metal characterized by
threefold charge order, bad metallic behavior and the emergence of high spin
local moments.
Our results naturally explain the anomalous charge-ordered metallic state
observed in the triangular layered compound AgNiO$_2$.
The potential relevance to other triangular transition metal oxides is
discussed.
\end{abstract}


\maketitle

Materials with  competing electronic interactions on triangular lattices are a
fertile ground for novel phenomena and original quantum phases, such as the
spin liquid behavior  \cite{Balents} observed in  organic
($\kappa$-(BEDT-TTF)$_2 $Cu$_2$(CN)$_3$,Me$_3$EtSb[Pd(dmit)$_2$]$_2$) and
inorganic  (Cs$_2$CuCl$_4$) quasi-two-dimensional materials.  Geometrical
frustration can play a similar role in charge ordering phenomena leading to
puzzling unconventional metallic and superconducting states.
Remarkable examples are  found in the quarter-filled organic salts
$\theta$-(BEDT-TTF)$_2$X \cite{Hotta12, Kanoda2013},    the layered cobaltates
Na$_x$CoO$_2$  \cite{Takada,CavaAlta,Julien2008,Alloul2012} and the transition metal
dichalcogenide 1T-TaS$_2$  \cite{Sipos}.

An interesting yet less explored member of this category is AgNiO$_2$, a
layered oxide compound with a triangular planar lattice structure, whose
properties reflect a rich interplay between magnetic, orbital and charge
degrees of freedom.  This system presents a robust 3-fold charge ordered phase,
which is stable above room temperature ($T_{\textrm{CO}}=365K$) and undergoes
magnetic ordering only at much lower temperatures, $T_N=20K$
\cite{Coldea07,Coldea08,Coldea11}.  Contrary to the common behavior observed in
oxides with Jahn-Teller active ions, the charge ordering in this material is
not associated to any structural distortion, indicative of a purely electronic
driving mechanism.  Furthermore, the ordering is partially frustrated by the
triangular lattice geometry, causing the electronic system to spontaneously
separate into localized magnetic moments, residing on a superlattice of
charge-rich Ni sites, and itinerant electrons moving on the honeycomb lattice
formed by the remaining charge poor Ni sites.  The material is therefore metallic
throughout the charge ordered phase, which contrasts with the situation in
non-frustrated perovskite nickelates  \cite{Mazin, Jackeli2008}, RNiO$_3$,
where charge order invariably leads to an insulating behavior. The high values
of the electrical resistivity  
and its anomalous temperature dependence in AgNiO$_2$ \cite{Sorgel05,Sorgel07,Shin,Wichainchai}, however, indicate bad metallic behavior,
  also supported by the observation of a large pseudogap in photoemission experiments
\cite{Kang07},  an anomalous Seebeck coefficient \cite{Wichainchai,Shin}, 
and a large specific heat coefficient  \cite{Coldea09}.

In this work, we analyze  a multiband microscopic model which takes explicit
account of electronic correlations to demonstrate that the emergence of charge
ordered phases with unconventional metallic properties is a natural outcome in
frustrated triangular oxides with both strong Coulomb interactions and Hund's
coupling.  Our results show that the combination of on-site and off-site Coulomb
repulsion and Hund exchange stabilizes of a robust {\it pinball
liquid} phase  \cite{Kaneko06,Hotta06,Miyazaki09,Cano10,Cano11,Merino13}, a quantum phase
where the charges spontaneously separate into coexisting localized (pins)
exhibiting Mott physics and  itinerant electrons (balls) moving on the remaining
honeycomb lattice.  We argue that the charge ordered metallic phase of
AgNiO$_2$ is a neat experimental realization of such pinball liquid, which
explains many experimental features such as the 3-fold ordering
pattern with strong charge disproportionation, the presence of large local
moments and the 'bad' metallic behavior.

{\it Two-orbital microscopic description.--}
\begin{figure}[h]
   \centering 
    \includegraphics[width=8cm]{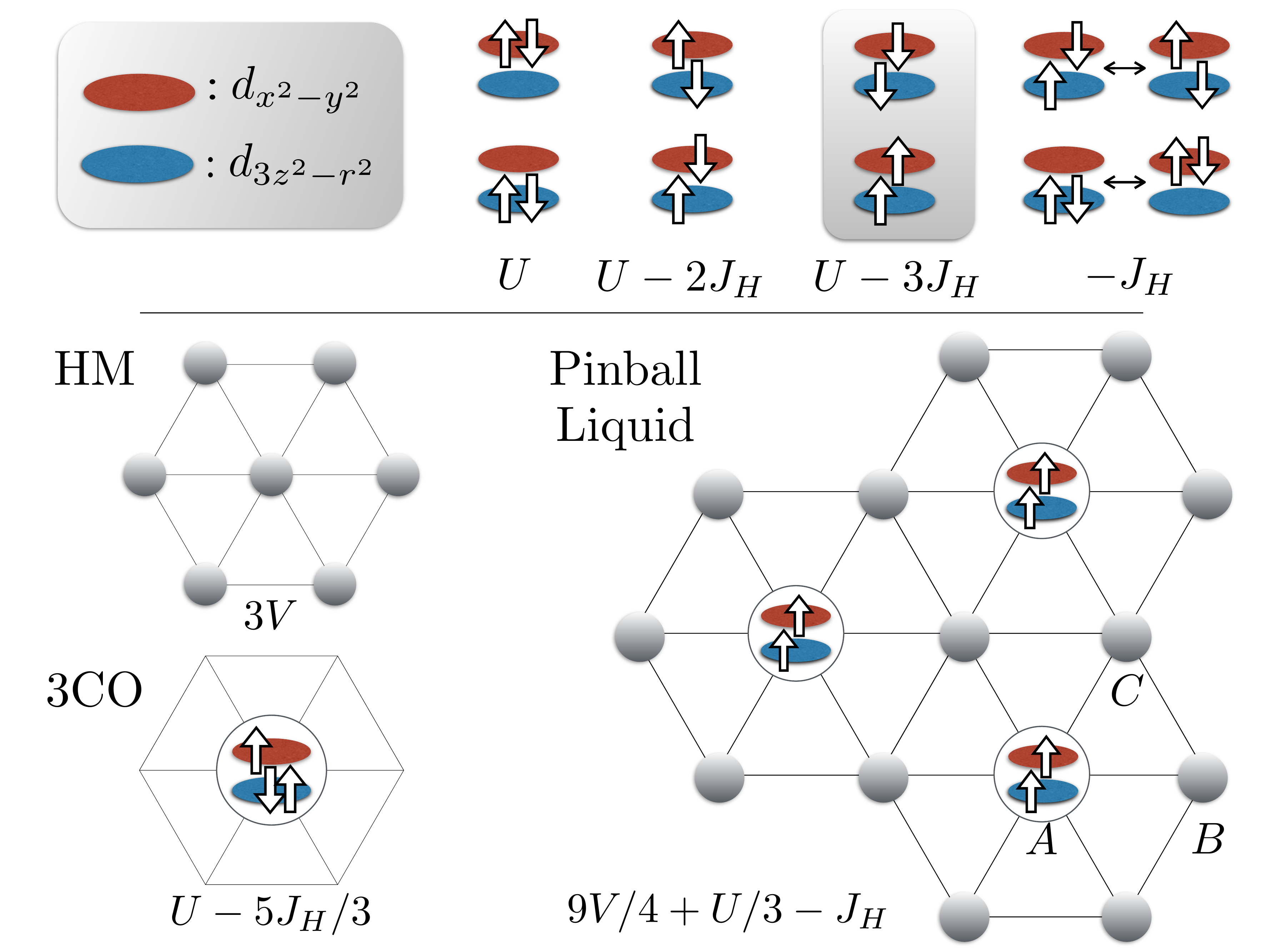}
   \caption{
   Top:  atomic processes described by the Hund interaction
$H_{\textrm{Hund}}$, with the corresponding coupling constants, highlighting
the high-spin configuration of lowest energy; the disks of different colors
represent the orbitals $d_{3z^2-r^2}$ and $d_{x^2-y^2}$; Bottom:  electronic
configurations in the homogeneous metal (HM), 3-fold charge ordered (3CO) and
pinball liquid phase (PL)   on the triangular lattice, with the corresponding
electrostatic energies.  Arrows represent localized moments (pins), gray disks
are the itinerant electrons (balls). }
 \label{fig:fig1}
 \end{figure}
In AgNiO$_2$ the d orbitals of Ni ($t^6_{2g}e^1{_g}$ configuration, formal
valence Ni$^{3+}$) split into an upper $e_g$ doublet occupied by one electron
and a completely filled lower $t_{2g}$ triplet, that are separated by a crystal
field gap of $\sim 2$ eV  \cite{VernayPRB04,Kang07,Sorgel07}.  Labeling by $\tau=1,2$ 
 the $e_g$ orbitals
$d_{3z^2-r^2}$ and $d_{x^2-y^2}$ and neglecting the low-lying $t_{2g}$
orbitals, we write the following two-orbital extended Hubbard model:
\begin{equation}
  \label{eq:H}
  H = -t \sum_{\langle ij\rangle} \sum_{\tau,\sigma} ( d_{i,\tau\sigma}^{\dagger} d_{j,\tau\sigma}+h.c.)
  + H_{\textrm{Hund}} + H_{\textrm{V}}. 
\end{equation}
The first term describes $e_g$ electrons  moving on the triangular lattice of
Ni  ions with transfer integrals $t$, at a density $n=1$ which nominally
corresponds to one quarter filling (one electron per two orbitals per site).
These interact on each atomic site via the Hund coupling, as described by the
standard Kanamori Hamiltonian \cite{Kanamori, Georgesrev}:
 \begin{eqnarray}
&& H_{\textrm{Hund}}=U\sum_{i,\tau} n_{i\tau\uparrow}n_{i\tau\downarrow}  
+(U-2J_H)\sum_{i,\tau\neq \tau^\prime} n_{i\tau\uparrow}n_{i\tau^\prime\downarrow}\nonumber\\ 
&& +(U-3J_H) \sum_{i,\tau<\tau^\prime,\sigma} n_{i\tau\sigma}n_{i\tau^\prime\sigma}  + \label{eq:ham_kanamori} \\
&& -J_H \sum_{i,\tau\neq \tau^\prime} 
\left(d^+_{i\tau\uparrow}d_{i\tau\downarrow}\,d^+_{i\tau^\prime\downarrow}d_{i\tau^\prime\uparrow} - 
d^+_{i\tau\uparrow}d^+_{i\tau\downarrow}\,d_{i\tau^\prime\downarrow}d_{i\tau^\prime\uparrow} \right). \nonumber 
\end{eqnarray}
This comprises intra-orbital and inter-orbital repulsion as well as pair
hopping and spin flip processes, as illustrated in Fig. \ref{fig:fig1}.  We
also consider a nearest neighbor Coulomb repulsion term $  H_{\textrm{V}} = V
\sum_{\langle ij\rangle}  n_i n_j$ as the driving force for charge
disproportionation, where $n_i=\sum_{\tau,\sigma} n_{i,\tau\sigma}$ is the
total density operator at site $i$, with
$n_{i,\tau\sigma}=d^+_{i,\tau\sigma}d_{i,\tau\sigma}$.

The competition between the different terms in the Hamiltonian Eq. (\ref{eq:H})
can be understood from the following electrostatic considerations.  For
sufficiently weak interactions, the system is a homogeneous metal (HM) with
$n_i=n=1$.  Because the interaction  energy $E_{\textrm{HM}}=3V$ of this
uniform configuration increases with $V$, a charge ordered configuration will
be preferred at large $V$ in order to minimize the electrostatic energy cost.
On the triangular lattice, this is achieved by ordering electrons on a three
sublattice structure (sublattices $A,B,C$) with $n_A=3$ electrons per site on
one sublattice and all other sites empty.
The  interaction energy of this threefold charge order (3CO), sketched in Fig. \ref{fig:fig1},  is  purely atomic,
$E_{\textrm{3CO}}=U-5J_H/3$  per site. It has no energy cost associated with
the off-site Coulomb repulsion, and is therefore favored at large $V$. 

From previous studies of the extended Hubbard model in the single band case
\cite{Cano10,Cano11,Merino13} it is known that an intermediate phase can be
stabilized between the 3CO and the homogeneous metal. In this phase, termed
{\it pinball liquid} (PL), part of the electron density of the charge-rich
sites ({\it pins}) spills out to the neighboring unoccupied sites ({\it balls})
in order to  reach a favorable compromise between local and non-local Coulomb
interactions.  The additional microscopic processes included in the present
multi-band case, which favor high spin configurations, play a key role in this
scenario: the maximum Hund's exchange energy is achieved in ions with a total
spin $1$ configuration, where precisely 2 electron spins are aligned (Fig.
\ref{fig:fig1}). Therefore, a  phase which maximizes the number of doubly
occupied sites will be naturally promoted for sufficiently large $J_H$,
stabilizing a pinball liquid state with $n_A=2$ on the charge rich sites
instead of $n_A=3$.

The key role of $J_H$ in stabilizing  the PL can be  assessed
quantitatively by comparing its energy, $E_{\textrm{PL}}=9V/4+U/3-J_H$, with
that of the 3CO and HM calculated previously.  The PL is favored when
$U_{c}^{(1)}<U<U_{c}^{(2)}$, with $U_{c}^{(1)} =J_H+(27/8)V+C_1$ and
$U_c^{(2)}=3J_H+(9/4)V+C_2$ (the constant terms arise from the kinetic energy
gain of mobile electrons in the PL and HM phases, which both scale
proportionally to $t$  \cite{Hotta06}).  This energetic argument predicts that
(i) the PL phase emerges  above a critical value of $J_H/U$; (ii) its area
spreads upon increasing $J_H/U$ and diverges for $J_H/U=1/3$.

{\it Phase diagram.--}

\begin{figure}[h]
   \centering 
    \includegraphics[width=8cm]{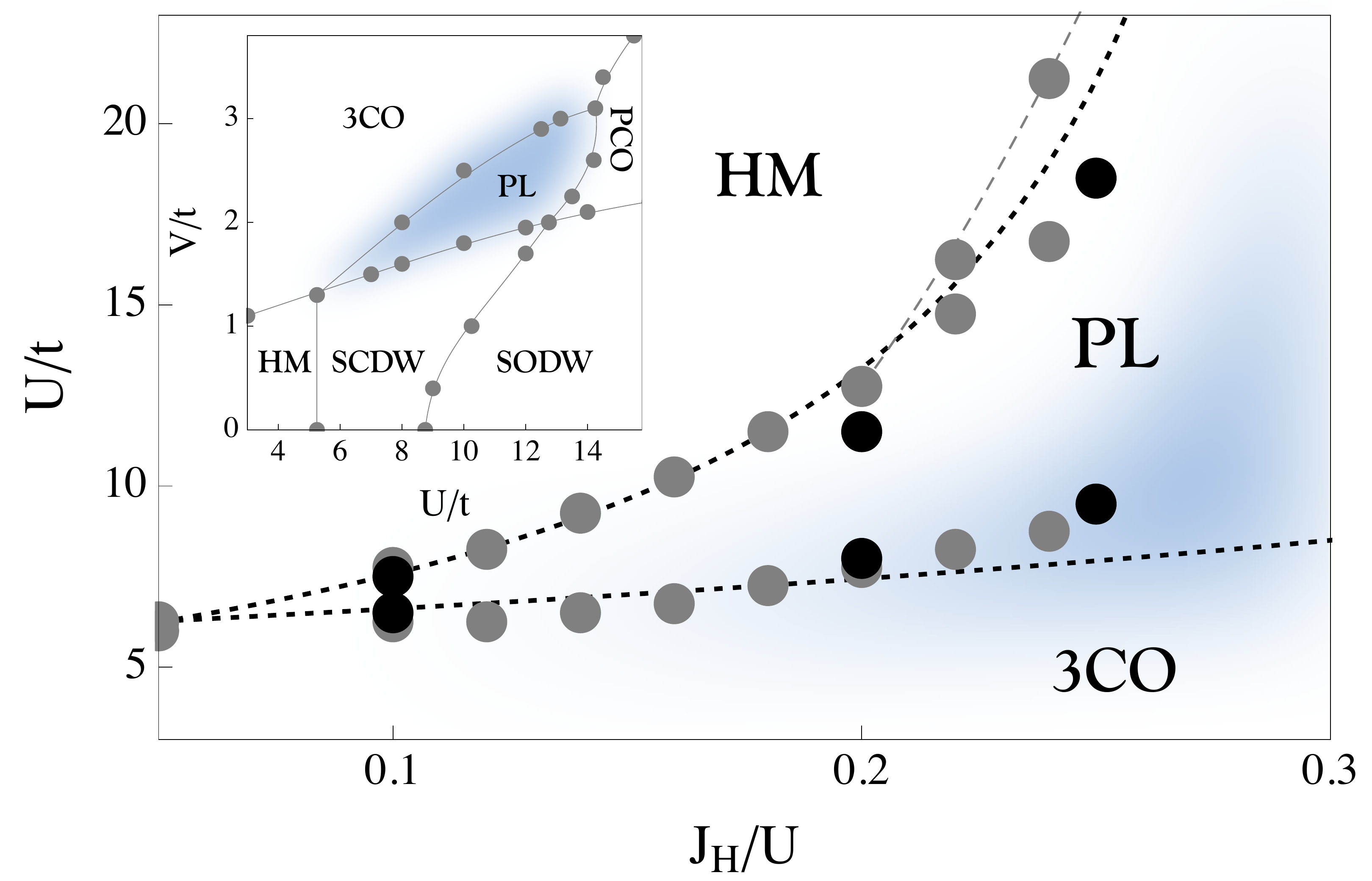}
   \caption{Phase diagram of the two-orbital extended Hubbard model on the
triangular lattice obtained from DMFT (black points) and UHF (gray points) for
a representative value $V/t=2$. The dotted lines are the phase boundaries
$U_{c}^{(1)}$ and $U_{c}^{(2)}$ given in the text.
The dashed line indicates charge order within the charge-poor sublattice as
found in the UHF solution.  Inset: UHF phase diagram in the $(U,V)$ plane, for
$J_H/U=0.2$.  }
 \label{fig:fig2}
 \end{figure}
 
We solve Eq. (\ref{eq:H})  by employing two complementary methods. We first
apply unrestricted Hartree-Fock (UHF) mean field theory allowing for solutions
breaking any symmetry, which guides us systematically through the whole phase
diagram.  To address
the effects of electron correlations that were neglected in previous
theoretical treatments \cite{Coldea07,Mazin,Uchigaito}, and which we
demonstrate here to be crucial in the region of experimental relevance, 
we then use single-site dynamical mean field theory (DMFT).  
This is expected to be particularly accurate in  systems with geometrical 
frustration or with large coordination, 
where the spatial range of non-local correlations is suppressed.  We
focus on solutions with three-fold  translational symmetry breaking,
restricting to paramagnetic phases  and ignoring the possible ordering on the
minority sublattice, which leaves us with 2 two-orbital impurity models
describing the charge-rich and charge-poor sites coupled only through the
hopping. Note that in the DMFT, the on-site correlations are treated exactly, while a Hartree decoupling is employed for the nearest-neighbor interaction V. The full DMFT self-consistent loop is evaluated using Lanczos
diagonalization until self-consistency. For
technical details see Refs. \cite{Cortes13} and \cite{Merino13} (second paragraph, left column).

We report in the main panel of Fig.  \ref{fig:fig2} the results obtained by
varying the ratio $J_H/U$ in the interval $(0.05-0.3)$ 
for an experimentally relevant value of the intersite
Coulomb repulsion $V/t=2$ \cite{notehop}.  As expected from the electrostatic
argument above, a prominent PL phase emerges in a broad region of the phase
diagram comprised between the homogeneous metal and the 3CO.  The boundaries of
the PL region, determined by the conditions $n_A=2$ (onset of PL) and $n_A=1$
(HM) are shown as points  (gray=UHF, black=DMFT) and closely follow the
analytical predictions $U_{c}^{(1)}$ and $U_{c}^{(2)}$,  drawn as dotted lines 
(here adjusted  by setting the constants $C_2= -C_1 = 0.8 t$).
The area covered by the PL spreads upon increasing the Hund coupling and
attains values of $U$ that are quite typical for transition metal oxides.  This
should be contrasted to the case where the Hund coupling is small or absent, in
which case the local Coulomb repulsion prevents any possibility of charge
ordering and a homogeneous metal is stabilized instead (left side of Fig.
\ref{fig:fig2}).

The physics of this model is even richer if we allow for more general
broken symmetry states, as presented in the inset of Fig.\ref{fig:fig2} for a
representative  value $J_H/U=0.2$.  The metal at low $V$ has further symmetry
breaking for sufficiently large $U$, corresponding to spin/charge density waves
(SCDW) and spin/orbital density waves (SODW). An additional  transition also
appears within the pinball phase at large $U$ and $V$, corresponding to the
ordering of the mobile electrons on the honeycomb lattice (Pinball Charge Order, PCO).  These results
will be discussed elsewhere  \cite{Clem}.

{\it Pinball liquid.--}
\begin{figure}
   \centering
   \includegraphics[width=8cm]{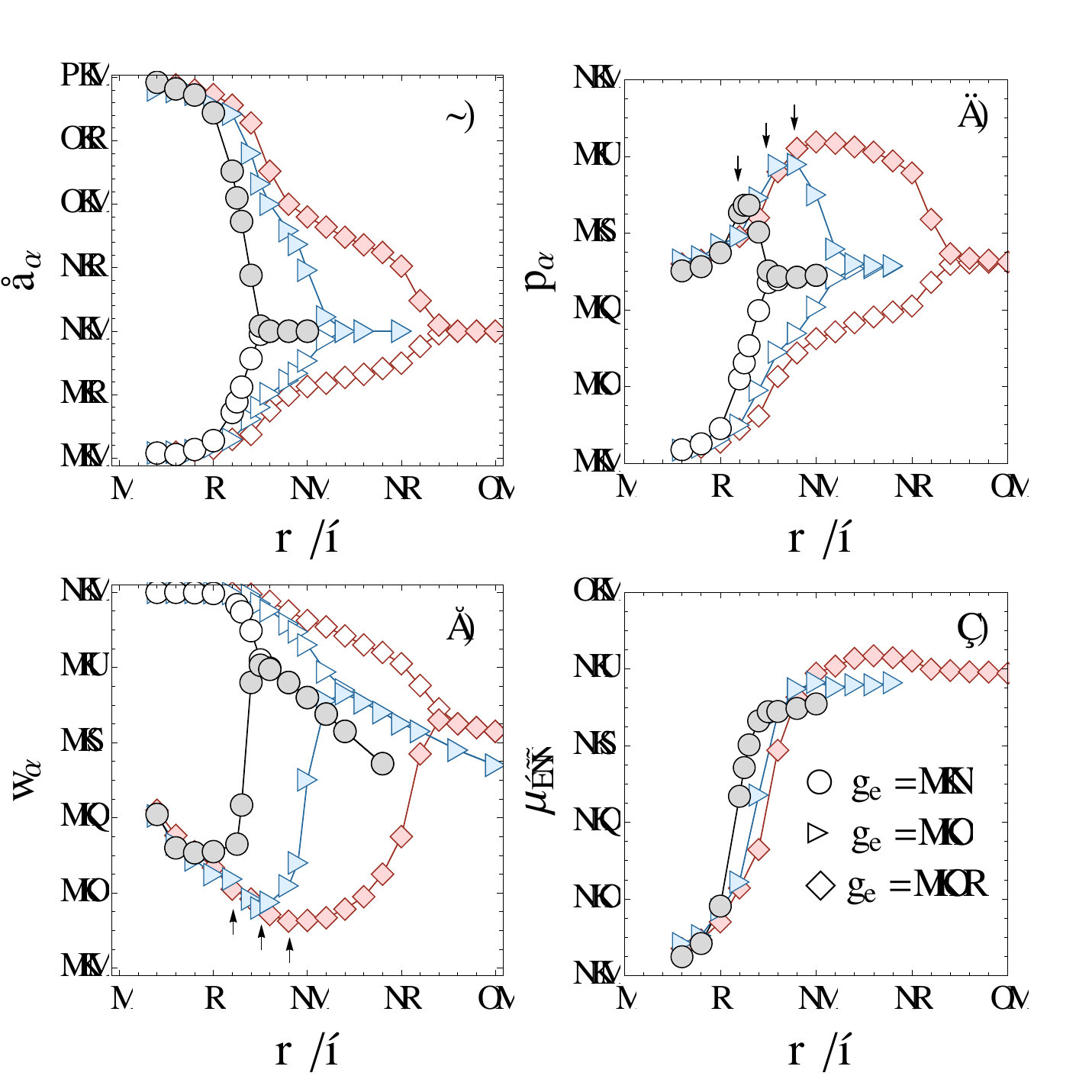}
	  \caption{(a) electronic density $n_\alpha$;  (b) magnetic moment $S_\alpha$ obtained from  
$\langle S^2_\alpha \rangle =S_\alpha(S_\alpha+1)$ and (c)  quasiparticle
renormalization $Z_\alpha$ on the
charge-rich ($\alpha=A$, filled symbols) and charge-poor sublattice
($\alpha=B,C$, open symbols), for $V/t=2$ and $J_H/U=0.1$ (circles), $0.2$
(triangles) and $0.25$ (diamonds);
(d) total effective Bohr magneton (see text).}
 \label{fig:fig3}
 \end{figure}
To  further characterize the PL phase, we show in Fig.  \ref{fig:fig3} several
physical properties obtained by DMFT at different values of the $J_H/U$ ratio.
The key quantity that controls the behavior of the system is the average
electron density in the different sublattices, shown in Fig. \ref{fig:fig3}(a).
Starting from the 3CO phase, the charge rich sublattice density is
progressively reduced with $U$ until it reaches $n_A=2$.  The onset of the PL
is signaled by a kink at this point, followed by a plateau which develops at
large $J_H/U$ extending all the way up to the HM phase.  To demonstrate that
such "lock-in" of the density is closely related to the existence of a
high-spin configuration on the pins, in Fig. \ref{fig:fig3}(b) we show the
value of the local magnetic moment evaluated for the same values of
the microscopic parameters. Closely following the behavior of the density, a
plateau is observed in the magnetic moment too, with a maximum in
correspondence of $n_A=2$ as expected (arrows).  Interestingly, upon increasing
$J_H$ the fluctuating magnetic moment takes large values approaching the ideal
limit $S_A=1$ ($S_A\approx 0.85$ at $U=9t$ and $J_H / U =0.25$), indicative of
strong local correlations.

The evolution of the quasiparticle weight, shown in  Fig. \ref{fig:fig3}(c),
reveals how in the presence of a sizable $J_H$, a large mass enhancement, $m^*/m_b=1/Z$, occurs
on the pins already for moderate values of $U\lesssim W$ (here $W=9t$ is the
bandwidth on the triangular lattice).  This  happens because the density on the
charge-rich sublattice  is locked around half-filling (two electrons in two
orbitals), which corresponds to the maximally correlated case in the presence
of Hund's coupling \cite{deMediciPRL11}.  Accordingly, the minimum of $Z_A$
coincides with the value where $n_A=2$, indicated by arrows.  At the same time,
the mass of the minority electrons remains close to the band value, owing to
their low concentration in the honeycomb lattice.  Fig. \ref{fig:fig3}(c) also
shows that the mass renormalization of the majority electrons in the PL phase
at intermediate $U$ is much stronger than that of the homogeneous metallic
phase at large $U$.  Upon reaching the HM phase, the quasiparticle weight jumps
back to a less correlated value.  It then gradually decreases with $U$ towards
the Mott transition expected at a value $U/t\sim 36$   \cite{deMediciPRB11}.
Note that within the HM phase the quasiparticle weight at a given $U$ is found
to increase with $J_H$ as expected for a two-orbital system with one electron
per site  \cite{deMediciPRB11}.

{\it Discussion.--} 
We now analyze the consequences of the present theoretical scenario, 
in connection with existing experiments on AgNiO$_2$.
To make a  quantitative comparison with the measured Curie-Weiss susceptibility
\cite{Sorgel05,Sorgel07,Coldea08}, we  report in Fig. \ref{fig:fig3}(d) the
effective  Bohr magneton,  $\mu_{\textrm{eff}}=g \mu_B \sqrt{  \langle
S^2_{\textrm{eff}} \rangle }$,  as obtained from the effective moment per site
evaluated in  DMFT: $\langle S_{\textrm{eff}}^2\rangle \approx {\langle
S_A^2\rangle+\langle S_B^2 \rangle + \langle S_C^2 \rangle  \over 3}$.  We see
that values much larger than the DFT-LDA estimates,  $\mu_{eff} = 1.3-1.5$
\cite{Coldea07} and quantitatively consistent with the experimental range of
results $\mu_{eff} = 1.81-1.96$, are naturally reached in the presence of
substantial electronic correlations in the PL, and remain high also in the HM
phase at larger $U$.

\begin{figure}[h]
   \centering
   \includegraphics[width=8cm]{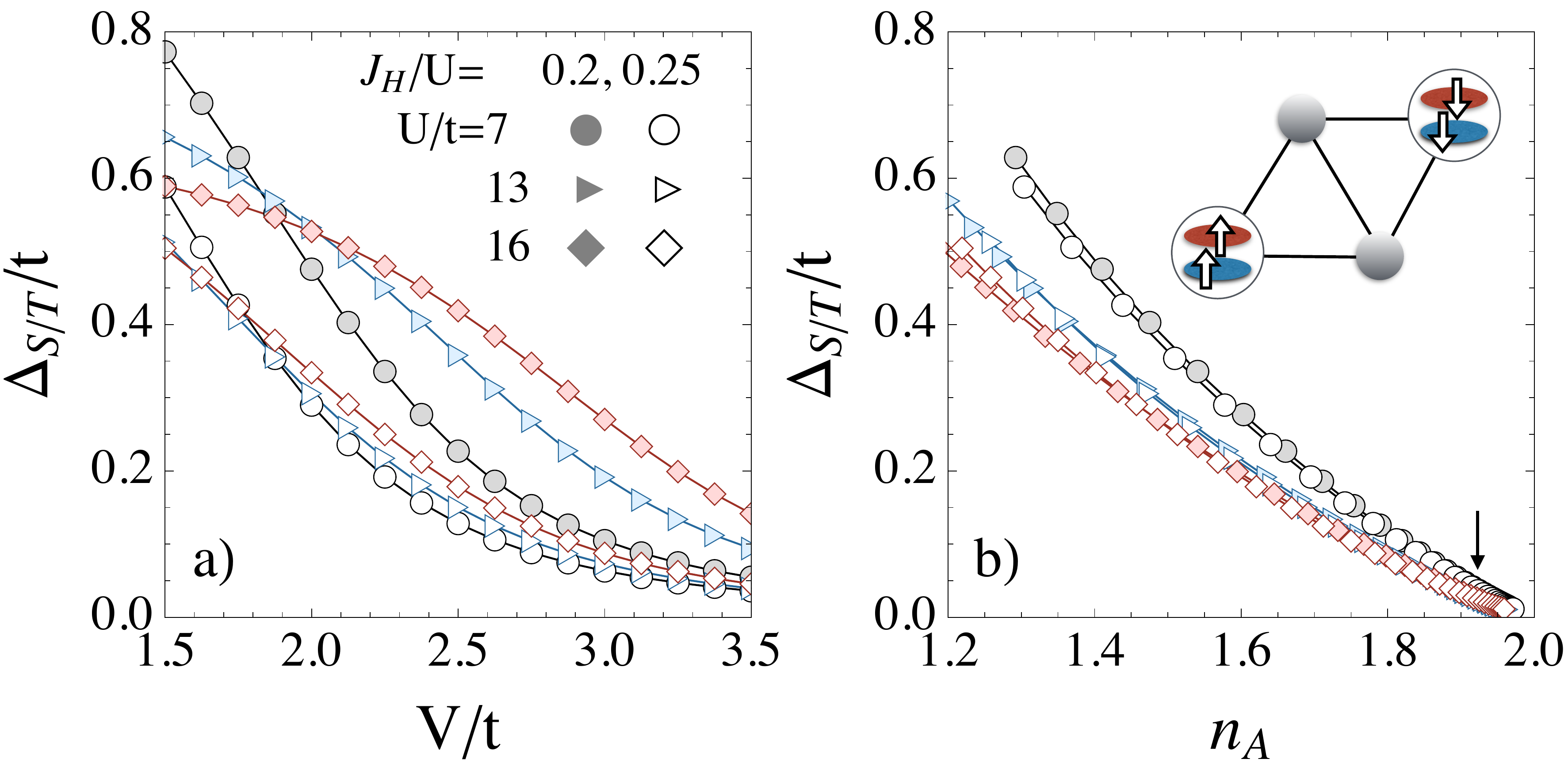}
   \caption{(a) Singlet-triplet gap for the model Eq. (1) on a 4-site cluster, in units of $t$.
(b) same, plotted vs.  the charge-rich sublattice density. } \label{fig:fig4}
 \end{figure}
 
Second, we discuss the origin of the 
magnetic ordering observed at low temperatures
\cite{Coldea07,Coldea08,Wheeler}. 
To assess the magnitude of the antiferromagnetic (AF)
coupling $J$ between nearest-neighboring local moments in
the PL mediated by virtual super-exchange processes via the charge-poor sites,
we  diagonalize Eq. (1)  on a minimal  4-site cluster with open boundary
conditions (illustrated in the inset of Fig. 4(b)), and calculate the singlet
triplet gap  $\Delta_{S/T}$ in the excitation spectrum, which coincides with
$J$ in the Heisenberg limit  \cite{noteorder}.  Fig. \ref{fig:fig4}(a) shows that $J$ strongly
decreases with $V$, and to a lesser extent also with $U$.  To highlight the
role played by the electron density on the pins, we  redraw the results as a
function of $n_A$ in Fig.  \ref{fig:fig4}(b). The data collapse in a narrow
region, which we use to estimate $n_A$ from the experimental value of $J$.
From the magnetic ordering temperature, $T_N=20K$ , and using $T_N\simeq 0.3 J$
based on the classical Monte Carlo simulations of Ref.  \cite{SeabraPRB11}, we
obtain $J\simeq 6 meV=0.03t$ for $t=0.2 eV$.  Such a low value of the magnetic
ordering temperature  implies that the system is very close  to the
integer filling $n_A=2$ (we estimate $n_A\gtrsim 1.9$, indicated by an arrow),
locating AgNiO$_2$ in the strongly
correlated pinball phase.

X-ray and neutron scattering experiments  do indicate substantial 3-fold charge
disproportionation \cite{Coldea08,Coldea11, Chung2008}, compatible with 
the emergence of large magnetic moments on the charge-rich sites \cite{Coldea07,Coldea08}.
In our scenario, an AF coupling between itinerant and localized species \cite{Merino13} 
leads to the screening of the pin local moments giving way to Fermi liquid behavior 
at low temperatures as in heavy fermions. Above the coherent-incoherent crossover temperature, $T^*$, quasiparticles
are destroyed due to the scattering of the itinerant carriers by the unscreened pin moments, which 
has several experimental manifestations.  The measured resistivity indeed displays typical Fermi liquid behavior $\rho \sim T^2 $ 
above the N\'eel temperature $T_N=20K$, albeit with anomalously large absolute 
values ($\gtrsim 1 m\Omega \ cm$) \cite{Sorgel05,Sorgel07,Wichainchai,Shin},
which crosses over to a (sub)linear 
$T$-dependence \cite{Sorgel05} at  temperatures above $T^*\simeq 150K$. 
At lower temperatures, the resistivity undergoes a sharp drop below $T_N$  which has 
been associated \cite{Coldea09,Coldea08} with the suppression of such scattering. 
The Seebeck coefficient increases linearly with temperature 
up to about 100 K as expected in a metal, but then it 
reaches a maximum around $T^*$  and changes  sign at 260 K \cite{Wichainchai,Shin}. 
A crossover in the  Curie-Weiss susceptibility
is also observed close to $T^*$ \cite{Sorgel07}.
Finally, the value of the specific heat coefficient, $\gamma=C_v/T$, within the AF phase suggests  
an appreciable mass enhancement, $m^*/m_b=2.6$  \cite{Coldea09}, intermediate between the values calculated 
for pins and balls in Fig.3(c).

{\it Outlook.--}
Previous works describing orbitally degenerate transition metal oxides with
quarter-filled bands have focused on models where frustration plays a minor
role, leading to charge ordered {\it insulating} states \cite{Jackeli2008,
Mazin}. Here, we have demonstrated that Coulomb induced charge ordering on
frustrated triangular lattices leads to the emergence of a robust {\it
metallic} pinball liquid phase stabilized by the Hund's coupling
acting on the d-orbital manifold. Such a phase presents characteristics qualitatively similar 
to heavy fermions and 
bad metallic behavior associated with the Kondo coupling between 
localized moments and itinerant carriers,
consistent with what is observed in the charge ordered metal AgNiO$_2$.
Optical conductivity experiments in this material could be used to    
observe the concomitant destruction of the Drude peak \cite{Merino00,Merino08,Kotliar14} above 
the coherent-incoherent crossover scale $T^*$.
Applying an  external pressure may 
destroy the magnetic order at a quantum critical point, 
giving way to a Fermi liquid state as observed in heavy fermion materials \cite{Si2008} 
and the nickel oxypnictide  CeNiAsO\cite{Georges2014}.
Other phases found here such
as the spin/orbital stripe states or the PCO phase with ordering of the ball
sites could also be realized, as is the case in adsorbates deposited on
metal surfaces \cite{Cortes13}.  Finally, a similar interplay of multi-orbital
physics, electronic correlations and charge ordering may occur on other
triangular compounds such as Ag$_2$NiO$_2$ \cite{Yoshida2006}, the
Ba$_3$B$^\prime$Ru$_2$O$_9$ ruthenates  \cite{Kimber}, the superconducting
cobaltates \cite{Takada,CavaAlta,Julien2008,Alloul2012} as well as other geometrically
frustrated lattices.

\begin{acknowledgments} Discussions with C. F\'evrier are gratefully
acknowledged. J. M.  acknowledges financial support from MINECO
(MAT2012-37263-C02-01).  This work is supported by the French National Research
Agency through Grants No.  ANR-12-JS04-0003-01 SUBRISSYME and No.
ANR-2010-BLANC-0406-0 NQPTP.
\end{acknowledgments}


\begin{thebibliography}{30}
\bibitem{noteorder}It is known that the ordering realized on the triangular lattice with
nearest-neighbor antiferromagnetic coupling is a canted anti-ferromagnet, while
the experimentally observed collinear stripe ordering requires a sizable
next-nearest-neighbor coupling  \cite{Jolicoeur,Seabra,SeabraPRB11,Coldea08,Wheeler}, which however
cannot be directly calculated within our method.
\end{thebibliography}
\end{document}